\documentclass[aps,prd,10pt,longbibliography,showkeys, breaklinks=true,  superscriptaddress, nofootinbib]{revtex4-2}
\usepackage[T1]{fontenc}
\usepackage[utf8]{inputenc}
\usepackage{color}
\usepackage{babel}
\usepackage{amsmath}
\usepackage{amssymb}
\usepackage{subfigure}
\usepackage{graphicx}
\usepackage{physics} 
\usepackage{dsfont}
\usepackage{hyperref}

\pdfoutput=1
\hypersetup{colorlinks=true,citecolor=red,linkcolor=blue,urlcolor=blue,filecolor= green, breaklinks=true}

\begin{document}

\title{Kerr-Newman outside a rotating de Sitter-type core: Different types of interior electromagnetic fields and currents}

\author{Marcos L. W. Basso}
\affiliation{Centro de Ci\^encias Naturais e Humanas, Universidade Federal do ABC, Avenida dos Estados 5001, Santo Andr\'e, S\~ao Paulo, 09210-580, Brazil.}
\affiliation{Departamento de Matemática Aplicada, Universidade Estadual de Campinas, 13083-859 Campinas, S\~ao Paulo, Brazil.}

\author{Vilson T. Zanchin}
\affiliation{Centro de Ci\^encias Naturais e Humanas, Universidade Federal do ABC, Avenida dos Estados 5001, Santo Andr\'e, S\~ao Paulo, 09210-580, Brazil.}

\begin{abstract}
We investigate the structure of interior electromagnetic fields compatible with the exterior Kerr–Newman geometry when regular interior geometries for describing regular black hole solutions are considered. Working within the G\"urses–G\"ursey framework and considering a rotating de Sitter–type geometry in the interior region, we analyze a particular class of electromagnetic fields aligned with the principal null directions and satisfying Maxwell electrodynamics. We show that, within such a class of interior electromagnetic fields, no well-behaved interior electromagnetic field can be supported in the absence of electromagnetic currents. Moreover, we show that for the same interior and exterior geometries, different types of interior electromagnetic fields and charge distributions, together with different charge densities on the rotating boundary shell, can be constructed while matching the same exterior Kerr–Newman electromagnetic field. This results highlight the non-uniqueness of the interior electromagnetic structure compatible with a given Kerr–Newman exterior. For a particular choice of interior electromagnetic fields we show that the de Sitter fluid can be interpreted as being electrically polarized by the rotating charged shell. In this case, the interior region contains no net electric charge, which is instead concentrated entirely on the shell, and the interior medium behaves as a perfect conductor with infinite conductivity. In a second choice for the electromagnetic field, we show that the polarized de Sitter fluid also contributes to the total electric charge, with the interior medium acting as a nonisotropic conductor. 
\end{abstract}


\maketitle
 
\section{Introduction}

Regular black holes (RBH) have been a subject of great interest in the literature.
Bardeen first introduced the concept of black holes with horizons but without curvature singularities, postulating a suitable mass function in a spherically symmetric metric,  marking the inception of regular black holes~\cite{Bardeen}. Such a mass function gives rise to an energy-momentum that was later associated with the presence of magnetic monopoles in nonlinear electrodynamics~\cite{Ayon}. At large radial coordinates, the geometry tends to the Reissner-Nordstr\"om metric, thus simulating the presence of an electric charge distribution throughout the spacetime. On the other hand, in the limit of small distances, close to the center, the energy-momentum tensor tends to a perfect fluid with a de Sitter equation of state, i.e., an isotropic fluid with energy density $\rho_m$ and negative pressure $p$ such that $p=-\rho_m$. 

Subsequent developments in the field have been largely based on Bardeen's original proposal, with a de Sitter-type core plus some electric charge therein or elsewhere, albeit with significant advancements in implementation and analysis~\cite{Frolov, Dymnikova92, Bronni01, Dymnikova04, Hayward06, Broni06, Zaslavskii2010, Lemos2011, Lemos16, Fan2016, Masa18, Simpson19}. 
In particular, the work by Lemos and Zanchin~\cite{Lemos2011} provides a brief review of the classification of regular black hole solutions through the type of junctions needed. If no junction exists, the solution remains continuous throughout the spacetime. When two spacetime regions are present, the solution incorporates a boundary surface connecting these regions. In more extreme scenarios, the solutions entail surface layers, such as thin shells, that link the two regions. Moreover, Ref.~\cite{Lemos2011} provides an interesting regular black hole solution, which consists of a Reissner-Nordstr\"om geometry outside a regular de Sitter-like core with an electrically charged shell at the boundary. This solution, that we dub the L\&Z electrically charged solution, is the static seed of the present work.

Although the search for RBH models has primarily focused on static configurations, efforts to explore their rotating counterparts have soon emerged. 
An interesting path to model such kind of rotating objects was built following the work by Newman and Janis~\cite{Newman65}. Furthermore, inspired by the Newman-Janis complex transformation~\cite{Newman65}, G\"urses and G\"ursey~\cite{Gurses} derived a stationary and axisymmetric metric in Boyer-Lindquist coordinates, laying out a framework for constructing rotating counterparts of static and spherically symmetric solutions. This approach holds promise for devising simple models for rotating RBH. Later, several avenues have been pursued to elucidate the source underlying the G\"urses-G\"ursey metric~\cite{Burinskii, Gondolo}, and to generalize and refine the Newman-Janis algorithm \cite{Burinskii, Gondolo, Drake, Keane, Bambi, Mustapha}.

Consequently, a substantial body of work has emerged on rotating regular objects~\cite{Spallucci, Modesto, Tosh14, Saa, Dymn15, Tosh17, Hernandez:2017fmg,Torres, Gosh, Mazza, Franzin, Masa22, Brustein, Dymn2023, Gosh23, Maeda, Ramon, Ramon23, Zhou23,Casadio:2023iqt, Dymn2024, Basso2025}. These endeavors stem from the recognition that while the Kerr geometry~\cite{Kerr67} provides a realistic description of the exterior region of black holes with angular momentum, it presents challenges regarding the interpretation of the black hole interior, including ring singularity and causality violations~\cite{Neil}. Substituting the problematic interior with a regular matter source, analogously to static RBH, offers a potential resolution for this issue. 

Following a similar strategy, one of the main goals of the present work consists in obtain and describe the rotating version of the L\&Z electrically charged solution. It is worth noting that in Ref.~\cite{Mustapha14}, a rotating de Sitter-like metric was derived using the modified version of the Newman-Janis algorithm without complexification. The resulting metric so obtained could describe a rotating version of the interior spacetime region of the L\&Z static solution.  However, in such a work the matching with an exterior geometry was not addressed, and the calculation of the interior electromagnetic field was omitted. In contrast, our work here presents explicit forms of all relevant quantities for the interior region and aims to establish a proper match with the exterior Kerr-Newman geometry~\cite{newman1965} together with the electromagnetic fields.

Interestingly, we show that it is possible to construct different types of interior electromagnetic fields together with different charge densities on the rotating shell that matches the exterior Kerr-Newman electromagnetic field. This result is in agreement with the investigation by Tiomno~\cite{Tiomno}, who found that a charged rotating oblate ellipsoid of revolution, which reproduces the exterior Kerr-Newman electromagnetic field in a flat spacetime, allows different types of interior electromagnetic fields with different electromagnetic media. 

Moreover, it should also be noted that several solutions describing charged rotating regular black holes have been derived within the framework of general relativity coupled to non-linear electrodynamics (see, e.g., \cite{Dymn15, Tosh17, Dymn2023}). However, considering a Lagrangian in the general form $\mathcal{L} = \mathcal{L}(F_{\mu \nu}F^{\mu \nu})$, it was shown in Ref.~\cite{Kubi22} that the standard Newman–Janis algorithm fails to produce rotating solutions in nonlinear electrodynamics, being consistent only with the linear Maxwell theory. In contrast, we show that the electromagnetic fields obtained in this work satisfy linear Maxwell electrodynamics in the G\"urses–G\"ursey geometry, which can be obtained from the Newman-Janis algorithm, in agreement with the results from Ref.~\cite{Kubi22}. 

The present work is organized as follows. The rotating G\"urses-G\"ursey geometry for charged systems is presented in Sec.~\ref{sec:ggmetric}, together with the particular class of electromagnetic fields considered in this work. In Sec.~\ref{sec:rotLZ} we construct the rotating version of the L\&Z electrically charged solution and analyze its main properties. We begin by reviewing the static L\&Z electrically charged solution. We then present the complete solution for the metric and electromagnetic fields, and perform the matching between interior and exterior regions with special interest in the boundary conditions of the electromagnetic fields. The different kinds of objects modeled by the complete solution are also briefly discussed in Sec.~\ref{sec:rotLZ}. Section \ref{sec:altEM} is dedicated to present and interpret an alternative electromagnetic field for the interior region of the rotating L\&Z electrically charged solution. Our final comments and conclusion are made in Sec.~\ref{sec:conc}. Appendix ~\ref{sec:basequ} contains a brief review of the fundamental equations for Einstein-Maxwell systems needed in our work, while Appendix ~\ref{sec:appMaxwell} provides some details showing that the chosen interior gauge fields presented in this work are actual solutions of the Maxwell equations in the G\"urses-G\"ursey spacetime.

\section{The electromagnetic field for Kerr-Newman-type metrics}
\label{sec:ggmetric}

\subsection{The G\"urses-G\"ursey metric}
\label{sec:ggmetric1}

As is well known, the complexification procedure devised by Newman and Janis~\cite{Newman65} generates new solutions of the Einstein equations and works for algebraically special metrics of the Kerr-Schild class.

Starting from a spherical seed metric of the form
\begin{align}
ds^2 = - \left(1-\frac{2{\mu}(r)}{{r}} + \frac{{\mathcal Q}^2(r)}{r^2}\right) dt^2 + \left(1-\frac{2\mu(r)}{{r}} + \frac{ {\mathcal Q}^2(r)}{r^2}\right)^{-1} dr^2 + r^2\left(d\theta^2 +\sin^2\theta\right), 
\label{eq:stmetric}
\end{align}
where $\mu(r)$ and ${\mathcal Q}(r)$ are the total mass and total electric charge inside a sphere of radius $r$, respectively.  The explicit form of the functions $\mu(r)$ and ${\mathcal Q}(r)$ will be presented in Sec.~\ref{sec:lemzan}. 

Then, the Newman-Janis and Gürses-Gürsey procedures yield a rotating and axially symmetric metric that, in Boyer-Lindquist coordinates $(t,\,r,\,\theta,\,\varphi)$, takes the form
\begin{equation}
\begin{aligned}
   ds^2  = & -\left(1 - \frac{2 r\, M(r) }{\Sigma(r,\theta)} \right) dt^2 + \frac{\Sigma(r,\theta)}{\Delta(r)} dr^2  - \frac{4 r\, M(r)  a \sin^2 \theta}{\Sigma(r,\theta)} dt\, d\varphi + \Sigma(r,\theta) d\theta^2 \\
    & + \left(r^2 + a^2 + \frac{2 r\, M(r) a^2 \sin^2 \theta}{\Sigma(r,\theta)} \right) \sin^2\theta\, d\varphi^2, \label{eq:ggmetric} 
\end{aligned}
\end{equation}
where $a$ is the rotation parameter, and the functions $M(r)$, $\Sigma(r,\theta)$, and $\Delta(r)$ are defined by 
\begin{align}
& M(r) = {\mu}(r) - \frac{ {\mathcal Q}^2(r)}{2r}, \ \ \  \Sigma(r,\theta)\equiv\Sigma = r^2+ a^2\cos^2\theta,\ \ \  \Delta(r)\equiv\Delta = r^2 + a^2 - 2r\, M(r), \label{eq:deltfun} 
\end{align}
respectively. 

Since the Newman–Janis complexification procedure and its formulation within the G\"urses–G\"ursey approach \cite{Gurses} are well established and widely used in the literature, we do not reproduce the details of the algorithm here. A comprehensive discussion of the method can be found in Refs.~\cite{Newman65,Gurses, Drake, Keane}, as well as in the subsequent literature of regular rotating black holes~\cite{Bambi, Masa22, Dymn2023}.

It is easy to see that metric \eqref{eq:ggmetric} reproduces the Kerr-Newman (KN) geometry in the case of constant $\mu(r)=m$ and constant ${\mathcal Q}(r)=q$, so that $M(r) = m -q^2/2r$. In the case of nonconstant $\mu(r)$ and ${\mathcal Q}(r)$, it describes a spacetime region fulfilled with charged matter. 
As discussed in our previous work \cite{BZ2024-paper1}, the metric with nonconstant $\mu(r)$ and ${\mathcal Q}(r)$ can be used to describe an interior spacetime region (${\cal M}_-)$ that may be joined to the KN metric, with the KN metric describing the exterior electrovacuum spacetime region (${\cal M}_+$).

\subsection{A particular class of electromagnetic fields}

With the aim of building interior geometries for the exterior Kerr–Newman (KN) metric, 
we study a particular class of electromagnetic fields in the Gürses–Gürsey geometry, together with their properties.

Let us first note that the electromagnetic gauge potential $\mathcal{A}_{\mu}$ in the KN spacetime is aligned with the principal null congruence 
$k_\mu dx^\mu  = dt - a \sin^2\theta \, d\phi$. Namely, in the Boyer-Lindquist coordinates, it assumes the particularly simple form,  $\mathcal{A}_\mu dx^\mu =  \dfrac{qr}{\Sigma} \left( dt - a \sin^2\theta \, d\phi\right)$, which explicitly exhibits its algebraic alignment of the electromagnetic field with the principal null directions of the geometry~\cite{Newman14}. In this work, we also adopt a Kerr–Newman–like electromagnetic gauge potential that is algebraically aligned with the principal null direction and can be obtained through the Newman–Janis procedure, as outlined in Refs.~\cite{Keane, Erbin}.

To implement this construction, we can follow the type-I complexification scheme proposed by Bambi and Modesto~\cite{Bambi}. The electromagnetic gauge potential of the static configuration is written as $\mathcal{A}_\mu dx^\mu = \phi(r) dt$, where $ \phi(r)$ is the electric potential, which is written in terms of the electric charge function $Q(r)$ as $\phi(r)=Q(r)/r+\,{\rm constant}$, with $Q(r)$ being an arbitrary function of $r$ only. 
Upon applying the procedure, after eliminating the radial component of $\mathcal{A}_{\mu}$ using the gauge freedom, the gauge potential assumes the form 
\begin{align}
\mathcal{A}_{\mu}dx^{\mu} = \frac{r\, Q(r)}{\Sigma(r,\theta)}\left( dt - a \sin^2\theta \, d\phi\right).    \label{eq:gaugeGG}
\end{align}
With this, the nontrivial components of a stationary and axially symmetric Faraday-Maxwell tensor in Boyer-Lindquist coordinates are $F_{rt}$, $F_{\theta t}$, $F_{r \varphi}$ and $F_{\theta \varphi}$. It is well established that the vanishing of the remaining components of the Faraday-Maxwell tensor is a consequence of the isometries inherent to all Kerr-like metrics, such as the one in Eq.~\eqref{eq:ggmetric}, particularly within electrovacuum regions. Furthermore, these spacetime symmetries impose an additional constraint on the non-vanishing components, expressed as
\begin{equation}
F_{\theta\varphi} = \frac{ F_{t\theta}\left[\left(r^2+a^2\right)\Sigma(r,\theta) -2\,\,r\, M(r)\left(F_{tr}\, a\, \sin^2\theta - F_{r\varphi}\right)\right] \sin^2\theta}{\left[\Sigma(r,\theta) - 2\,r\, M(r)\right]F_{r\varphi}+ 2\, r\, M(r)\, a \sin^2\theta F_{tr}}.\label{eq:constraintF}
\end{equation}

The gauge potential \eqref{eq:gaugeGG} also leads to specific algebraic relations among the non-vanishing components of the Faraday-Maxwell tensor $F_{\mu \nu}$. In Boyer–Lindquist coordinates, these relations read as follows,
\begin{align}
    F_{r\varphi} = - a \sin^2 \theta F_{rt}, \ a F_{\theta \varphi} = -\left(r^2+a^2\right) F_{\theta t}. \label{eq:a1}
\end{align}
Notice that such relations satisfy the constraint \eqref{eq:constraintF}, and are exactly the same as in Kerr-Newman electrovacuum solution.

\subsubsection{The electric and magnetic fields and the total electric charge}

It is helpful to decompose the Faraday-Maxwell tensor $F_{\mu\nu}$ into its electric and magnetic components relative to an observer comoving with the source of the Gürses–Gürsey geometry. As is well known, the electric and magnetic fields measured by an observer comoving with the fluid are given by
\begin{equation}
 E_{\mu} = F_{\mu \nu} u^{\nu}, \ \ \  B_{\mu} = \frac{1}{2} \epsilon_{\nu \mu \alpha \beta} u^{\nu} F^{\alpha \beta}, \label{eq:EBfields}
\end{equation}
where $u^\mu$ is the four-velocity of the comoving observer with the fluid, which, in the present case, is given by the first element of the Carter tetrad $e^\mu_0$ (see Eq.~\eqref{eq:rtetrad} in Appendix \ref{sec:basequ}) and $\epsilon_{\beta \mu \nu \alpha}$ is the usual four-dimensional Levi-Civita tensor. 
After the definitions in \eqref{eq:EBfields}, the Faraday-Maxwell decomposes in the form
\begin{align}
  F_{\mu \nu}= u_{\mu} E_{\nu} - u_{\nu} E_{\mu}  + \epsilon_{\mu \nu \alpha} B^{\alpha},
\end{align}
where $\epsilon_{\mu \nu \alpha} = \epsilon_{\beta \mu \nu \alpha}u^{\beta}$ is the three-dimensional Levi-Civita tensor defined in the subspace orthogonal to the four-velocity of the comoving observer. 

By using Eq.~\eqref{eq:a1} and the projections given by Eq.~\eqref{eq:EBfields}, a straightforward calculation shows that the only nonzero components of the electric and magnetic fields are
\begin{align}
    & E_r =\!\sqrt{\frac{\Sigma}{\pm \Delta}} F_{rt},  \ \ \ B_r = - \sqrt{\frac{\Sigma}{\pm \Delta}} \frac{F_{\theta t}}{a \sin \theta}. \label{eq:ErBr} 
\end{align} 
This decomposition admits the usual physical interpretation in terms of electric and magnetic fields as measured by the Carter tetrad only in regions where the associated four-velocity remains timelike, specifically, outside the event horizon and inside the Cauchy horizon. A noteworthy feature here is that, as perceived by the comoving observer, the electric and the magnetic fields are parallel to each other. This was already noticed by Tiomno~\cite{Tiomno} and Lynden-Bell~\cite{Lynden2004} for the flat spacetime limit of the Kerr-Newman electromagnetic field. In addition, Gair and Lynden-Bell~\cite{Lynden2007} also verified that, considering electromagnetic fields in Carter separable spacetimes, there exists an observer for every spheroidal surface such that the electric and magnetic fields are parallel to each other.

Finally, another important quantity for our analysis is the total electric charge within a surface of constant coordinate $r$. Given the hypersurface $\mathcal{S}_t$ defined by $t  = \text{constant}$, the total electric charge of the distribution is defined through the volume integral
\begin{align}
    \mathfrak{q} = - \int_{\mathcal{S}_t} J^{\mu} \eta_{\mu} d \mathcal{V} = - \frac{1}{4\pi} \int_{\mathcal{S}_t} \left(\nabla_{\nu} F^{\mu \nu}\right) \eta_{\mu} d \mathcal{V}, 
\end{align}
where $\eta_{\mu}$ is a unitary timelike vector orthogonal to the surface $\mathcal{S}_t$, and $d \mathcal{V} = \sqrt{h}\, dr\, d \theta\, d \varphi$, with $h$ being the determinant of the induced metric in $\mathcal{S}_t$.
Then, the total electric charge contained within a spheroid of radius $r$ is given by
\begin{align}
    \mathfrak{q}(r) & = \frac{1}{2} \int_0^{r} dr' \int_0^\pi d\theta \,\partial_{\nu}\left(\sqrt{-g}F^{t\nu}\right) = \frac{1}{2}  \int_0^r dr' \int_0^\pi d\theta  \partial_{r'} \left[\left(r'^2 + a^2\right) \sin \theta F_{rt} \right]\nonumber \\
    & = \frac{1}{2} \int_0^\pi\Big[\big(r'^2 + a^2\big) \sin \theta F_{rt} \Big]_0^{r}d\theta, \label{eq:totalq}
\end{align}
where we used the fact that $\sqrt{-g} F^{tr} = -(r^2 + a^2) \sin \theta F_{tr}$, and that the $\theta$-component on the summation when integrated over the polar angle does not contribute. In order to get such relations we assumed that the integrand is well-behaved everywhere inside the spheroid. In the present work, Eq.~\eqref{eq:totalq} will be used only in Secs.~\ref{sec:rotLZ} and~\ref{sec:altEM}, where we deal with electromagnetic field tensors that are, in fact, well-behaved everywhere in the interior region, including the ring $S^1\!\!:\!(r = 0,\, \theta = \pi/2)$.

\subsubsection{On the nonexistence of well-behaved Kerr-Newman-like electromagnetic field with vanishing current density}
\label{sec:imp}
Here we show that no well-behaved Kerr-Newman-like electromagnetic fields are possible in the interior region without the presence of an electromagnetic current density.

Let us start by writing the Maxwell equations [see Eqs.~\eqref{eq:Maxw} and~\eqref{eq:Max2} in Appendix \ref{sec:basequ}] for the G\"urses-G\"ursey metric~\eqref{eq:ggmetric} with vanishing current density, $J^\mu=0$, i.e., 
\begin{equation}
\begin{aligned}
    & \partial_r\left[\left(r^2 + a^2\right)\sin \theta F_{rt}\right] + \partial_{\theta}\left[\sin \theta F_{\theta t}\right] = 0, \qquad \ \  \partial_r \left[\csc\theta F_{r \varphi} \right] + \partial_{\theta} \left[\frac{\csc\theta}{r^2 + a^2} F_{\theta \varphi} \right] = 0,\\
    & \partial_{r} F_{\theta t} -  \partial_{\theta} F_{rt} = 0, \qquad  \  \ \ \ \ \ \ \ \ \ \ \ \ \ \ \ \ \ \ \ \ \ \ \ \ \ \ \ \ \  \ \partial_{r} F_{\theta \varphi} - \partial_{\theta} F_{r \varphi} = 0, 
\end{aligned} \label{eq:max0}
\end{equation}
where $\partial_{\mu} \equiv \partial/\partial x^{\mu}$.

After condition \eqref{eq:a1}, the set of the Maxwell equations \eqref{eq:max0} can be rewritten in terms of the components $F_{rt}$ and $F_{\theta t}$ only, i.e.,
\begin{align}
    & \partial_r\left[\left(r^2 +a^2\right)\sin \theta F_{rt}\right] + \partial_{\theta}\left[\sin \theta F_{\theta t}\right] = 0,\qquad  \ \ \ \  \partial_r \left[a \sin \theta F_{r t} \right] + \partial_{\theta} \left[\frac{\csc\theta}{a} F_{\theta t} \right] = 0,
    \label{eq:mx6} \\
    & \partial_{r} F_{\theta t} -  \partial_{\theta} F_{rt} = 0, \qquad \qquad \qquad \ \ \ \ \ \ \ \ \ \ \ \ \ \ \ \ \ \ \ \ \ \partial_{r}\left[\left(r^2+a^2\right) F_{\theta t}\right] - \partial_{\theta}\left[a^2 \sin^2 \theta F_{r t}\right] = 0. \label{eq:mx8}
\end{align}

Here we closely follow the work of Dymnikova and Galaktionov~\cite{Dymn23}. By manipulating the set of Eqs.~\eqref{eq:mx6}, it is possible to obtain the following set of equations
\begin{align}
    & \frac{\partial}{\partial \theta}\left[\frac{\partial}{\partial r}\left(\frac{\Sigma}{r \sin \theta} F_{\theta t} \right) + \frac{2 F_{\theta t}}{\sin \theta} \right] = 0, \qquad \ \ \ F_{rt} = \frac{1}{2a^2 r \sin \theta} \frac{\partial}{\partial \theta} \left(\frac{\Sigma}{\sin \theta} F_{\theta t} \right). \label{eq:auxeq1}
\end{align}

The integration of the first relation of Eq.~\eqref{eq:auxeq1} in variable $\theta$ furnishes the relation
\begin{align}
\frac{\partial}{\partial r}\left(\frac{\Sigma}{r \sin \theta} F_{\theta t} \right) + \frac{2 F_{\theta t}}{\sin \theta} = \Phi(r), \label{eq:auxeq2}   
\end{align}
where $\Phi(r)$ is an arbitrary integration function that depends on the coordinate $r$ alone. 
Moreover, by introducing the function $V_{\theta t} = \Sigma\, F_{\theta t}/(r \sin \theta)$, Eq.~\eqref{eq:auxeq2} may be recast into the following differential equation
\begin{align}
    \frac{\partial}{\partial r} V_{\theta t} + \frac{2r}{\Sigma} V_{\theta t} = \Phi(r),
\end{align}
which gives us
\begin{align}
     V_{\theta t} = \Sigma^{-1} \left( \int^r\! \Sigma \Phi(r) dr + \Psi_0(\theta) \right), 
\end{align}
where $\Psi_0(\theta)$ is another integration function that depends on the coordinate $\theta$ alone. From these results, we can fully describe the components $F_{\theta t}$ and $F_{r t}$ in terms of the integration functions $\Phi(r)$ and $\Psi(\theta) = \sin \theta \Psi_0(\theta)$, i.e.,
\begin{align}
 \!\!\!\!\!   F_{\theta t} & = \frac{r }{\Sigma^2} \Psi(\theta) + \frac{r \sin \theta}{\Sigma^2} \int^r\! \Sigma \Phi(r) dr, \label{eq:Fot} \\
\!\!\!\!\!  F_{rt} & =  \frac{\csc^2 \theta}{2a^2 \Sigma^2} \Big[\Sigma \frac{d}{d\theta}\Psi(\theta) - \Psi(\theta) \tan \theta\left(\Sigma - 2 a^2 \sin^2 \theta \right) \Big] 
   + \frac{\cos \theta}{\Sigma^2} \left(\int^r\! \Phi(r) r^2 dr - r^2 \int^r\! \Phi(r) dr \right), \label{eq:Frt} 
\end{align}
where we also used the second relation in \eqref{eq:auxeq1} to get the last expression for $F_{rt}$.

The integration functions $\Phi(r)$ and $\Psi(\theta)$ can be determined by the two remaining Maxwell equations, i.e., the relations in Eq.~\eqref{eq:mx8}, which can be put into the form
\begin{equation}
\begin{aligned}
    & \frac{\partial}{\partial r} \left[\frac{\partial}{\partial \theta}\left(\frac{\Sigma}{\sin 2 \theta} F_{\theta t} \right) - a^2 F_{\theta t} \right] = 0, \qquad \ \ F_{rt} = \frac{1}{a^2 \sin 2 \theta} \frac{\partial}{\partial r}\left[\left(r^2+a^2\right) F_{\theta t} \right]. \label{eq:max3}
\end{aligned} 
\end{equation}
Hence, by substituting the expressions \eqref{eq:Fot} and \eqref{eq:Frt} into the relations in \eqref{eq:max3}, after a tedious but straightforward calculation, we get the following equations for $\Phi(r)$ and $\Psi(\theta)$,
\begin{equation}
\begin{aligned}
&     \sin \theta\,\frac{d}{dr}\! \left[\Phi(r)\,r^2\right] = 0,\\
&  \frac{d}{d\theta}\Psi(\theta)  + \left(\tan \theta - \cot \theta \right) \Psi(\theta) - \sin \theta \tan \theta \Big( r \Sigma \Phi(r)  -  \int^r\!\! \Phi(r) r^2 dr \Big) - a^2 \sin^2 \theta \cos \theta \int^r\! \Phi(r) dr = 0. \label{eq:pp1}
\end{aligned}
\end{equation}

The first equation in \eqref{eq:pp1} yields
\begin{equation} 
     \Phi(r) = k_1\, r^{-2}, \label{eq:phi}   
\end{equation}
with $k_1$ being an arbitrary constant.
Upon inserting $\Phi(r)$ from Eq. \eqref{eq:phi} into the second equation of \eqref{eq:pp1}, the r-dependent terms cancel. Consequently, the equation simplifies to a form that is easily integrated to give 
\begin{equation} 
     \Psi(\theta) = k_2\sin 2\theta, \label{eq:psi}
\end{equation}
with $k_2$ being another integration constant.

Hence, substituting these functions into Eqs.~\eqref{eq:Frt} and \eqref{eq:Fot}, we get the general expression for the electromagnetic fields with no current density in the G\"urses-G\"ursey geometry as
\begin{equation}
\begin{aligned}
& F_{rt} =  \frac{2 k_1 r \cos \theta}{\Sigma^2} -  \frac{k_2\left(r^2 - a^2 \cos^2 \theta\right)}{a^2 \Sigma^2}, \qquad\ \ \ F_{\theta t} = \frac{k_1 \sin \theta \left(r^2 - a^2 \cos^2 \theta\right)}{\Sigma^2} +  \frac{k_2 r \sin 2 \theta}{\Sigma^2}.
\end{aligned}    
\end{equation}

We first notice that the well-known Kerr-Newman electromagnetic field is recovered by choosing $k_1 = 0$ and $k_2 = - qa^2$. The second important fact to note is that, for any nonzero arbitrary constants $k_1$ and $k_2$, the electromagnetic field is not well-behaved at the ring ($r = 0,\,\theta = \pi/2$). Indeed,
\begin{equation}
\begin{aligned}
    & \lim_{r \to 0} \left(\lim_{\theta \to \pi/2} F_{rt} \right) = -\lim_{r \to 0} \frac{k_2}{a^2 r^2} \to - \infty, \qquad \ \ \  \lim_{r \to 0} \left(\lim_{\theta \to \pi/2} F_{\theta t} \right) = \lim_{r \to 0} \frac{k_1}{ r^2} \to \infty.
\end{aligned}    
\end{equation}

In conclusion, the electromagnetic field is well-behaved only if it vanishes completely in the interior, which occurs by taking $k_1 = 0$ and $k_2 = 0$.

\subsubsection{No solution with zero electric and magnetic fields in the central region}

As demonstrated, the only well-behaved Kerr-Newman-like electromagnetic field lacking a current density in the central region of a regular G\"urses-G\"ursey metric~\eqref{eq:ggmetric} is one that vanishes entirely within that region. Notably, this result is independent of the mass function $M(r)$ and thus holds for any valid choice of $M(r)$ within the Gürses–Gürsey geometry \eqref{eq:ggmetric}.

In order to better understand the result obtained in Sec.~\ref{sec:imp}, let us then consider the matching of the vanishing interior electromagnetic field with the exterior Kerr-Newman field on a surface of constant radial coordinate $r=r_0$. The matching of the normal components of the electric and magnetic fields $E_\mu$ and $B_\mu$, as defined in Eq.~\eqref{eq:EBfields} at $r=r_0$ implies that $E_{r}\big|_{ext} - E_{r}\big|_{int} = 4\pi \sigma_e$ and $B_{r}\big|_{ext} - B_{r}\big|_{int} = 0$, where $\sigma_e$ is some surface electric charge density. 
Hence, if the inner components vanish, i.e., if $E_{r}\big|_{int}=0=B_{r}\big|_{int}$, the boundary condition for $B_r$ cannot be satisfied, unless the exterior magnetic field also vanishes. This means that, in the metric \eqref{eq:ggmetric}, the Maxwell equations with a surface charged shell are satisfied only in the static case, for which the rotation parameter $a$ is zero, and the magnetic field also vanishes everywhere throughout the spacetime [see Eq.~\eqref{eq:ErBr}]. 

Therefore, in order to find a regular solution that holds in the central region of the geometry, more specifically in the Kerr-like ring ($r\to 0$, $\theta\to \pi/2)$, it is necessary to consider well-behaved electromagnetic fields with some current density as a source. The result we have just found motivates us to investigate such a scenario, which we will do in the following sections.

\section{Kerr-Newman outside a rotating de Sitter-type core}
\label{sec:rotLZ}

\subsection{Preliminary remarks}

Following the Gürsey-Gürses algorithm to generate a stationary metric from a static seed (summarized in Sec.~\ref{sec:ggmetric1}), we derive a rotating counterpart to the electrically charged solution proposed by Lemos and Zanchin \cite{Lemos2011}. As reviewed in the following section, the static seed geometry comprises an external Reissner-Nordström spacetime joined to an internal de Sitter core by an electrically charged shell. After deriving the rotating metric, we investigate the interior electromagnetic fields required to satisfy matching conditions with the exterior Kerr-Newman geometry.

\subsection{The static (seed) metric}

\label{sec:lemzan}

For completeness, here we present a brief summary of the static L\&Z \cite{Lemos2011} electrically charged solution. 
This solution corresponds to a static spherically symmetric spacetime composed of two disjoint regions that join together at a spherical surface $\mathcal{B}_r$ which, in Schwarzschild-like coordinates, coincides with the surface of constant radial coordinate, i.e., $\mathcal{B}_r\!\!:r = r_0=$ constant. The interior region ($r< r_0$) contains an uncharged perfect fluid obeying a de Sitter equation of state,  while the exterior region ($r > r_0$) is described by the electrovacuum Reissner-Nordstr\"om (RN) solution. The surface $\mathcal{B}_r$ bears a total electric charge $q$ uniformly distributed on it. 

The metric employed in Ref.~\cite{Lemos2011} can be conveniently expressed using Schwarzschild-like coordinates $(t,\, r,\, \theta,\, \varphi)$ as in \eqref{eq:stmetric}, where $\mu(r)$ is the total mass inside a sphere of radius $r$, and is then given by
\begin{equation} 
\mu(r) = \frac{r^3}{2R^2} \left[1 - \Theta(r-r_0)\right] + m \,\Theta(r - r_0),
\label{eq:mass1}
\end{equation}
with $\Theta(x)$ being the Heaviside function, such that $\Theta(x) = 1$ for $x \ge 0$ and $\Theta(x) = 0$ for $x <0 $.
The parameter $R$ admits a physical interpretation similar to that in the Schwarzschild interior solution and in the analogous interior solutions for charged matter distributions previously discussed in the literature. In such constructions, $R^{-2}$ is proportional to the energy density, and the term $3R^{-2}$ effectively plays the role of a cosmological constant in the Einstein–Maxwell equations. Accordingly, $R$ can be interpreted as a de Sitter radius~\cite{Lemos2011}.

Meanwhile,  ${\mathcal Q}(r)$ is the total electric charge inside a sphere of radius $r$ and is given by
\begin{equation}
{\mathcal Q}(r) = q\,\Theta(r - r_0).
\label{eq:charge1}
\end{equation}

The smooth matching conditions between the two spacetime metrics yield two constraints for the parameters of the models. Namely,
\begin{align}
    & \frac{m}{R} = 2 \frac{r_0^3}{R^3}, \ \ \  \frac{q}{R} = \sqrt{3}\frac{r_0^2}{R^2}. \label{eq:bound2}
\end{align}
 Notice that, with conditions  the metric function $F(r)= 1 -2\mu(r)/r + {\mathcal Q}^2(r)/r^2$ and its derivative $dF(r)/dr$ are continuous functions across the boundary surface $\mathcal{B}_r$.

The interior region contains a perfect fluid that satisfies a de Sitter type equation of state, i.e., $p(r) =-\rho_m(r)$, while the exterior region is the Reissner-Nordstr\"om (RN) electrovacuum spacetime. The fluid quantities are given by
\begin{eqnarray}
8\pi \rho_{\rm m}(r) = -8\pi p(r)= \left(\dfrac{3}{R^2} -\dfrac{ {\mathcal Q}^2(r)}{r^4}\right) \Theta(r_0 - r).\label{rhom-pressure}
\end{eqnarray}

For future reference, it is also convenient to write here the complete solution for the electromagnetic quantities of the L\&Z solutions.
The charge density profile may be written as
\begin{equation} 
 \rho_e(r) = \dfrac{q}{4\pi\, r_0^2}\,\delta(r-r_0),\label{eq:chargeden}
\end{equation}
where $\delta(r-r_0)$ stands for the Dirac delta function, indicating that the total charge $q$ is uniformly distributed on the boundary surface $\mathcal{B}_r$. In this case, the electric potential $\phi(r)$ is given by 
\begin{equation} \label{eq:stphi}
\phi(r) = \frac{q}{r_0} \left[1 - \Theta(r-r_0)\right] + \dfrac{q}{r} \Theta (r - r_0), 
\end{equation}
while the electric field $E(r)$ presents a jump at $r=r_0$, 
vanishing in the interior region, i.e., 
\begin{equation} \label{eq:stE(r)}
E(r) = \dfrac{q}{r^2} \Theta(r - r_0).
\end{equation}
The discontinuity in the normal components of the electromagnetic field at the boundary surface $\mathcal{B}_r$ is directly related to the surface charge density, defined as $ \sigma_e = {q}\big/({4\pi\, r_0^2})$. This relationship is derived from the standard electromagnetic field matching conditions~\cite{Lemos2011}.

\subsection{The interior rotating solution}
\label{sec:intEMa}

The interior geometry is characterized by the G\"urses-G\"ursey metric~\eqref{eq:ggmetric} with the mass function $M(r)$ given by
\begin{equation}
M(r) = \mu(r) = \frac{r^3}{2R^2}. \label{eq:m-(r)}
\end{equation}
This is a solution to the Einstein equations with a particularly simple energy-momentum tensor, as we show below (see also Appendix~\ref{sec:basequ} for more details). 

Once the interior metric is established, we turn our attention to the interior electromagnetic field. In order to define the gauge potential for the interior region, let us recall that, in the interior region of the L\&Z static solution, the electric potential is constant and given by $\phi(r) = q/r_0$, and the net charge inside any sphere of constant radius $r < r_0$ is zero. Hence,
a possible form of the electromagnetic gauge potential $\mathcal{A}_{\mu}$, which is consistent with the geometry \eqref{eq:ggmetric}, is constant along the coordinate $r$, and also leads to a vanishing net charge inside any spheroid of constant radius $r < r_0$, is given by 
\begin{align}
    \mathcal{A}_{\mu} = \frac{q r_0}{\Sigma_0}\left(\delta_{\mu}^{\ t} - a \sin^2 \theta \delta_{\mu}^{\ \varphi}\right), \label{eq:rotpot1} 
\end{align}
with $\Sigma_0 = r_0^2 + a^2 \cos^2 \theta$. Notice that the gauge potential \eqref{eq:rotpot1} depends only on the polar coordinate $\theta$. 
In this case, from Eq.~\eqref{eq:rotpot1}, we have the following Faraday-Maxwell tensor, 
\begin{equation}
F_{rt} = 0= F_{r \varphi}, \label{eq:Frt0}
\end{equation}
and the nontrivial components of the Faraday-Maxwell tensor are
\begin{equation}
     F_{\theta t} = - \frac{qr_0 a^2 \sin 2 \theta}{\Sigma_0^2}, \qquad \ \ F_{\theta \varphi} = -\frac{\left(r_0^2+a^2\right)}{a} F_{\theta t}. \label{eq:electenLZ}
\end{equation}

It is easy to verify that the two nontrivial components of $F_{\mu\nu}$ vanish in the limit $a \to 0$, recovering the static L\&Z solution.
It can also be seen from the result \eqref{eq:electenLZ} that the Faraday-Maxwell tensor is well-behaved in the entire interior region, being finite everywhere inside the fluid.  In particular, all components of $F_{\mu\nu}$ vanish for $\theta=0$ and for $\theta = \pi/2$ for all $r < r_0$, including the ring $S^1\!\!:\!(r = 0,\, \theta = \pi/2)$. 

Interestingly, we notice that once the electric field (as measured by the comoving observer with the fluid) vanishes, $E_{\mu} = F_{\mu \nu} u^{\nu} = 0$, the fluid behaves like a perfect conductor, in agreement with the results found by Tiomno~\cite{Tiomno} for the interior material that reproduces the exterior Kerr-Newman electromagnetic field in flat spacetime.
In fact, from Eq.~\eqref{eq:ErBr} and the vanishing of $F_{rt}$ in the interior region, it follows that the radial component of the electric field $E_r$ vanishes, while the radial component of the magnetic field is given by
\begin{align}
B_r  = \sqrt{\frac{\Sigma}{\pm \Delta}}  \frac{2 q\, r_0\, a \cos  \theta}{\Sigma_0^2}.     
\end{align}

The current density in the interior region that results from the Faraday-Maxwell field \eqref{eq:electenLZ} is given by
\begin{align}
     J^t = & \frac{q r_0 a^2}{2\pi \Sigma \,\Sigma_0^3} \left[\left(r_0^2 - 3 a^2 \cos^2 \theta\right)\sin^2 \theta - 2\Sigma_0 \cos^2 \theta \right], \ \ \ J^{\varphi} =  \frac{q r_0 a}{2 \pi \Sigma\, \Sigma_0^3}\left(\frac{r_0^2 + a^2}{r^2 + a^2}\right)\left(r_0^2 - 3 a^2 \cos^2 \theta\right), \label{eq:Jphi-}
\end{align}
with the other components being identically zero. As seen from Eq.~\eqref{eq:Jphi-}, the two nontrivial components of the current density are finite everywhere in the interior region, with the exception of the ring $S^1$, where both components diverge. 

\begin{figure}
    \centering
    \includegraphics[width=.45\columnwidth]{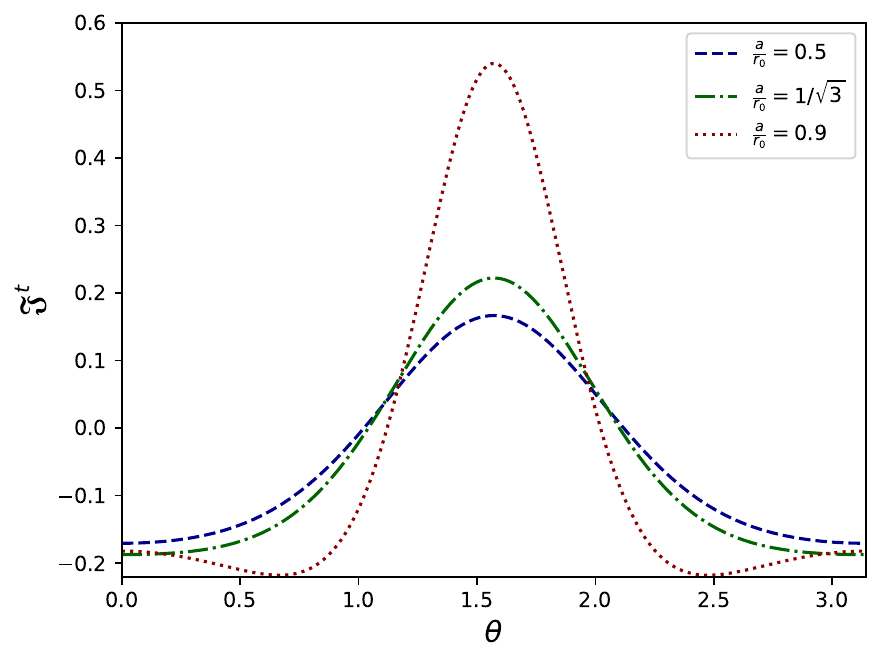}
    \caption{The normalized $t$-component of the current density at the boundary surface $r=r_0$, defined as $\mathfrak{J}^t = J^t/(4 \pi q r_0^3 / 3)$, as a function of $\theta$ for different values of $a/r_0$. $J^t$ is given in Eq.~\eqref{eq:Jphi-}. }
    \label{fig:Jtq}
\end{figure}

Figure~\ref{fig:Jtq} shows the behavior of the $t$-component of the current density $J^t$ on the boundary surface $r = r_0$, as a function of $\theta$ for some values of the rotation parameter $a/r_0$. As can be seen in the figure, $J^t$ is negative near the poles ($\theta = 0, \pi$), while it becomes positive near the equator ($\theta = \pi/2$), indicating that the medium in that region is electrically polarized. This behavior of $J^t$ is observed for any $r < r_0$. Moreover, once $J^t \to 0$ for $a \to 0$, we can infer that this polarization is induced by the rotation of the de Sitter-like fluid. 

As expected, the electromagnetic field does not vanish in the interior region. The rotating spheroidal charged shell induces internal electric and magnetic fields, which in turn generate a current density. This allows the matter content within that region to be interpreted as a charged rotating de Sitter-like fluid. However, because the induced electromagnetic field satisfies the condition $F_{rt}=0$, it follows from Eq.~\eqref{eq:totalq} that the net electric charge within any spheroid of radius $r<r0$ vanishes, i.e., 
\begin{equation}
    \mathfrak{q}(r)=0, \ \ r<r_0. \label{eq:inner-q}
\end{equation}
As we show below, the net charge is distributed on the boundary spheroidal shell located at $r=r_0$. 

We may refine the analysis of the charge distribution within the fluid by considering a comoving observer in that region.
The four velocity $u^\mu$ of such a comoving observer with the fluid coincides with the first element of the Carter tetrad \eqref{eq:rtetrad}, i.e., $u^{\mu} = e_{0}^{\ \mu}$. Then, the current density $J^\mu$ may be decomposed as
\begin{align}
    J^{\mu} = \rho_e u^{\mu} + j^{\mu}, \label{eq:jmu-}
\end{align}
where
\begin{equation}
\begin{aligned}
    &  \rho_{e}=  - J^{\mu}u_{\mu} = \sqrt{\frac{\pm \Delta}{\Sigma}} \left(J^t - a \sin^2 \theta J^{\varphi} \right),  \qquad j^{\mu} = \left(u^{\mu}u_{\nu} + \delta^{\mu}_{\ \nu} \right)J^{\nu} =\left(e^{3}_{\ \nu}J^{\nu} \right) e_{3}^{\ \mu},  \label{eq:Jdecomp}
\end{aligned}
\end{equation}
with $ e_3^{\ \mu}$ standing for the fourth component of the Carter tetrad \eqref{eq:rtetrad}. The quantities defined in \eqref{eq:Jdecomp} are interpreted as the charge density and the conduction current density as measured by a comoving observer, respectively. In the
present case, they reduce to
\begin{align}
    \rho_{e} = & \sqrt{\frac{\pm \Delta}{\Sigma}} \frac{q r_0 a^2 }{2 \pi \Sigma \,\Sigma_0^3} \left[\frac{r^2 - r_0^2}{r^2 + a^2}\left(r_0^2 - 3 a^2 \cos^2 \theta\right) \sin^2 \theta  - 2\Sigma_0 \cos^2 \theta \right], \label{eq:rhoe-} \\
    j^{\mu} = & \frac{qr_0 a(r_0^2 - a^2 \cos^2 \theta)}{2 \pi \Sigma \,\Sigma_0^2} \frac{a \sin \theta}{\sqrt{\Sigma}} e_3^{\ \mu}. \label{eq:conduction-}
\end{align} 
Thus, the nonzero spatial current can exist alongside a zero electric field, provided the medium has infinite conductivity~\cite{Tsagas}.

It is worth mentioning once again that a different gauge potential yields a different result also for the net charge inside the fluid.  An alternative electromagnetic field for the rotating L\&Z electrically charged solution is analyzed in Sec.~\ref{sec:altEM}.

\subsection{The exterior rotating solution: Kerr-Newman}

The exterior solution is characterized by the G\"urses-G\"ursey metric~\eqref{eq:ggmetric} with the mass function $M(r)$ given by 
\begin{equation}
M(r) = m - \frac{q^2}{2r}. \label{eq:m(r)-ext}
\end{equation}
Therefore, the exterior solution is the well-known Kerr-Newman solution~\cite{newman1965}, with the electromagnetic gauge potential given by
\begin{equation}
 \mathcal{A}_{\mu} = \frac{q\,r}{\Sigma}\left(\delta_{\mu}^{\ t} - a \sin^2 \theta \delta_{\mu}^{\ \varphi}\right), \label{eq:rotpot+}
\end{equation}
from which the nontrivial components for the Faraday-Maxwell tensor are obtained
\begin{align}
    & F_{rt} = \frac{q\left(r^2 - a^2 \cos^2 \theta\right) }{\Sigma^2}, \qquad \ \ F_{r\varphi} = - a \sin^2 \theta F_{ rt}, \qquad \ \  F_{\theta t} = - \frac{q\, r\, a^2 \sin 2 \theta}{\Sigma^2}, \qquad \ \ F_{\theta \varphi} = -\frac{\left(r^2+a^2\right)}{a} F_{\theta t}. \label{eq:electen+}  
\end{align}

\subsection{Matching conditions}
\label{sec:match-metric}

Moving on to the matching conditions, the interior region $\mathcal{M}_-$ corresponds to $r < r_0$, while the exterior region $\mathcal{M}_+$ corresponds to $r_0 < r < \infty$. We first address the metric junction conditions using the Darmois-Israel (DI) approach~\cite{Israel66}, followed by the matching conditions of the electromagnetic field.

The boundary between the two spacetime regions is the surface $\mathcal{B}_r$ defined by $r = r_0 = \text{constant}$.
Then, let $\xi^a\equiv (\tau, \,\theta, \,\varphi)$ be the coordinates in $\mathcal{B}_r$, and $x^{\mu}_{\pm}= \big(t_\pm$, $r_\pm$, $\theta_\pm$, $\varphi_\pm\big)$ be the coordinates of regions $\mathcal{M}_+$ and $\mathcal{M}_-$, respectively. 
Notice that, due to the symmetry of the spacetime and of the boundary surface, and since we are considering smooth boundary conditions for the geometric quantities, the coordinates of the two spacetime regions and the boundary may be identified, i.e., $t_-=t_+ = \tau$, $r_-=r_+$, $\theta_-=\theta_+=\theta$, and $\varphi_-=\varphi_+=\varphi$. Hence, from now on we drop the $\pm$ indexes.
The DI approach deals with the first and second fundamental forms on ${\cal B}_r$, $h_{ab}$ and $K_{ab}$, respectively. These geometric objects are defined in terms of four-dimensional geometric objects via the well-known relations $h_{ab} = \varepsilon_a^{\ \mu} \varepsilon_b^{\ \nu}g_{\mu \nu}$ and $K_{ab} = \varepsilon_a^{\ \mu} \varepsilon_b^{\ \nu} \nabla_{\nu} n_{\mu}$, where $\varepsilon_{a}^{\ \mu} = \partial x^{\mu} / \partial \xi^a = \delta_a^{\mu}$. Here, $n_{\mu} = \sqrt{g_{rr}} \delta_{\mu}^{\ r}$ denotes the unit normal vector on the surface ${\cal B}_r$.

The assumption of a smooth transition between the two spacetime regions across the boundary $\mathcal{B}_r$, i.e., $[h_{ab}] = 0$ and $[K_{ab}] = 0$, furnishes
\begin{equation}
    \frac{1}{R^2} = \frac{1}{r_0^3} \left(2 m - \frac{q^2}{r_0} \right),\qquad \  \frac{1}{R^2} = - \frac{1}{r_0^3} \left( m - \frac{q^2}{r_0} \right), \label{eq:match2}
\end{equation}
which implies the boundary condition given by Eq.~\eqref{eq:bound2}, being the same constraints found in the static solution.

Similarly to the static L\&Z solution, the Faraday-Maxwell tensor is not continuous across the boundary, requiring a charged shell. The matching conditions for the gauge potential and the electromagnetic field tensor are thus given by
\begin{equation}
\left[\mathcal{A}_a\right] = \left[\varepsilon_a^{\ \mu} \mathcal{A}_\mu \right] = 0,
\ \ \ 
\left[F_{ab}\right] =\left[ \varepsilon_a^{\ \mu} \varepsilon_b^{\ \nu}F_{\mu \nu}\right] =0, \ \ \ \left[F_{an}\right] = \left[\varepsilon_a^{\ \mu} n^{\nu}F_{\mu \nu}\right]= 4 \pi \sigma_e u_a, \label{eq:matchF}
\end{equation}
where $\sigma_e$ is the surface charge density and $u^a$ is the velocity of the comoving observer at the boundary shell, which is given by the $e_0^\mu$ element of the Carter tetrad~\eqref{eq:rtetrad}.

It is easy to verify that the first two conditions in Eq.~\eqref{eq:matchF} are trivially satisfied, while the last condition gives us the charge density of the shell as measured by an observer comoving with the boundary, i.e.,
\begin{align}
    \sigma_e= \frac{q \left(r_0^2 - a^2 \cos^2 \theta\right)}{4 \pi \Sigma_0^2}. \label{eq:rotchdlz}
\end{align}
Let us notice that the result for the charge density of the spherical shell in the static solution \eqref{eq:chargeden} is recovered from \eqref{eq:rotchdlz} in the limit $a=0$. In addition, we can see that the charge density of the shell can change its sign depending on the values of $r_0/a$, being positive for $\cos \theta < r_0/a$ and negative for $\cos \theta > r_0/a$, which implies that the rotating shell is also electrically polarized.

The integration of the charge density \eqref{eq:rotchdlz} along the surface $r=r_0$ gives the result $q$. Moreover, as already mentioned above [see Eq.~\eqref{eq:inner-q}], the integration of the component $J^t$ of the current density given in Eq.~\eqref{eq:Jphi-} over the whole interior region gives zero, confirming that the total charge of the distribution is exactly $q$, as expected. Therefore, we can think of the rotating de Sitter fluid in the interior region as being electrically polarized by rotation.

\subsection{Brief comment on curvature regularity}
 
It is pertinent to note that, as shown by Torres~\cite{Torres} and Maeda~\cite{Maeda}, the G\"urses-G\"ursey metric with mass functions of the form $M(r) = m_0 r^{3 + \alpha}$ for $\alpha \geq 0$, upon extension to $r<0$, exhibits a ring-like conical singularity at the ring $S^1$, rather than a scalar polynomial curvature singularity like the Kerr and Kerr-Newman spacetimes. 
On the other hand, as also shown by Torres~\cite{Ramon23}, if the extension to $r<0$ is not carried out, the ring $S^1$ is also devoid of a conical singularity. 
In the present case, for the interior region, where the ring is located, the mass function is given by Eq.~\eqref{eq:m-(r)}, i.e., the mass function is of the form $m_0 r^{3 + \alpha}$ with $\alpha=0$ implying that the ring $S^1$ does not present a curvature singularity. However, as noted in several investigations~\cite{Spallucci, Modesto, Torres, Maeda, Masa22, Basso2025}, even when assuming finite values at the ring, the curvature scalars are not well defined for $\alpha = 0$. This is because the curvature scalars depend on the path taken when approaching the ring. This phenomenon also occurs in the present case. For a more detailed discussion of this issue, see Ref.~\cite{Basso2025}.

\subsection{The energy-momentum tensor}

\subsubsection{Interior energy-momentum tensor}

Using Carter's orthonormal tetrad as defined in Eq.~\eqref{eq:rtetrad}, the total EMT acquires a diagonal structure $T_{\mu\nu}e^\mu_a e^\nu_b = {\rm diag}(\varrho,\, \mathfrak p_1,$ $\, \mathfrak p_2,\, \mathfrak p_3)$. The eigenvalues $\varrho$ and $\mathfrak p_1$ are interpreted as the total energy density and the effective radial pressure, respectively. The eigenvalues $\mathfrak p_2$ and $\mathfrak p_3$ are the effective tangential pressures. Equation~\eqref{eq:tpressa}, together with the mass function given by Eq.~\eqref{eq:m-(r)}, leads to the following expressions for the eigenvalues
\begin{equation}
8\pi  \varrho = - 8\pi \mathfrak p_{1} = \frac{3}{R^2}\frac{r^4} {\Sigma^2}, \qquad \ \ 8\pi  \mathfrak p_{2} = 8\pi \mathfrak p_{3} = \frac{3}{ R^2}\frac{r^4} {\Sigma^2}\left( 1- \frac{2\Sigma}{r^2}\right). \label{eq:rhopresslz-}
\end{equation}
Notice that in the limit $r\to 0$, the energy density and pressures present the same behavior as the curvature scalars analyzed in the previous section.

The contribution of the electromagnetic field to the EMT is obtained from Eqs.~\eqref{eq:emtEMa} and \eqref{eq:electenLZ}, in which it is enough to calculate the energy density as measured by the Carter's frame, namely,
\begin{align} \label{eq:rhopressEMlz-}
    &  8\pi  \mathfrak p_{em 2}  = 8\pi \mathfrak p_{em 3}= -8\pi \mathfrak p_{em 1} = 8\pi \varrho_{em} = \frac{4q^2r_0^2a^2\cos^2\theta}{ \Sigma_0^4}. 
\end{align}
In this case, we can see that the energy density and pressures depend only on the coordinate $\theta$ and vanish for $\theta = \pi/2$.

The contribution of matter to energy density and pressures can be obtained by subtracting $E_{\mu\nu}$ from $T_{\mu\nu}$. In Carter's frame, the eigenvalues of the matter EMT are given by
\begin{equation}
  8\pi \varrho_{m} =  - 8\pi \mathfrak p_{m 1} = \frac{3m}{2r_0^3}\left(\frac{r^4}{\Sigma^2} - \frac{4r_0^6a^2\cos^2\theta}{\Sigma^4_0} \right), \qquad \ 8\pi \mathfrak p_{m 2} = 8\pi \mathfrak p_{m 3}= 8\pi \varrho_{m} - \frac{3 mr^2}{r_0^3 \Sigma}, \label{rhopressmater-}
\end{equation}
which represents a non-isotropic fluid. We see that, different from the static case, the fluid quantities do not vanish at $r = r_0$.

\subsubsection{Exterior energy-momentum tensor}

Finally, using the relation in Eq.~\eqref{eq:tpressa}, together with the mass function $M(r) = m - q^2/2r$, we obtain the explicit expressions for the eigenvalues valid in the exterior region $\mathcal {M}_+$. That is,
\begin{equation}
   \mathfrak p_{2} = 8\pi \, \mathfrak p_{3} = -8\pi\,\mathfrak p_{1} =   8\pi \varrho = \frac{q^2}{\Sigma^2},  \label{eq:rhopresslz+}
\end{equation}
that is due to the presence of the electromagnetic vacuum field, i.e., $\varrho = \varrho_{em}$.

By comparing the relations in \eqref{eq:rhopresslz+} with the relations in \eqref{eq:rhopressEMlz-} it is seen that the EMT of the electromagnetic field alone is not continuous across the boundary $r=r_0$ due to the charged shell. However, it should be noted that the energy density and the radial pressure of the total (effective) EMT are continuous through the boundary shell. This can be easily seen by comparing the relations for the energy density and pressures in Eq.~\eqref{eq:rhopresslz+} with the corresponding relations in Eq.~\eqref{eq:rhopresslz-}, and using the constraints $Rq = \sqrt{3} r_0^2$ and $R^2 m = 2 r_0^3$.

\subsubsection{Brief comment on energy conditions}

In~\cite{Torres}, Torres demonstrated that if the mass function $M(r)$ of the G\"urses-G\"ursey metric can be expanded as a Taylor polynomial series around $r = 0$, then the weak energy condition should be violated near $r = 0$. Furthermore, Maeda~\cite{Maeda} extended Torres findings by showing that if the mass function $M(r)$ can be expanded around the locus $r = 0$, with $\theta \neq \pi/2$, as $M(r) \approx m_0 r^{3 + \alpha}$ with $\alpha \ge 0$, then all energy conditions are violated in the vicinity of $r = 0$ for $m_0 > 0$, while the null energy condition and the strong energy condition hold for $m_0 < 0$. Consequently, given that, in our case, the mass function is given by Eq.~\eqref{eq:m-(r)}, resulting in $m_0 = R^{-2}/2 > 0$, it becomes evident that all energy conditions are infringed around the region $r = 0$, $\theta \neq \pi/2$.

\subsection{On the horizons and ergosurfaces of the rotating geometry}

The constraints $Rq = \sqrt{3} r_0^2$ and $R^2 m = 2 r_0^3$, which are obtained from the matching conditions \eqref{eq:match2}, when substituted into Eq.~\eqref{eq:deltfun} allow us to obtain the horizons radii $r_h$ as the real and positive roots of the following polynomial equations,
\begin{equation}
 \Delta_-(r) = r^2 + a^2 - \frac{m}{2r_0^3}r^4 = 0, \qquad  \  \Delta_+(r) = r^2 + a^2 - 2mr + \frac{3 m r_0}{2} = 0, \label{eq:horieq6}
\end{equation}
with the supplemental conditions that the horizons are given by the roots of $ \Delta_-(r)$ if $r_h \le r_0$, while, for $r_h \ge r_0$, the horizons are given by the roots $ \Delta_+(r)$.
These polynomial equations can be easily solved in terms of radicals, and the solutions are 
\begin{equation}
  r_{k-} = r_0 \sqrt{\frac{r_0}{m}\left(1 + (-1)^k \sqrt{1 +  \frac{2m}{r_0} \frac{a^2}{r_0^2}}\right)}, \qquad \ \ r_{k+} = m \left(1 + (-1)^k \sqrt{1 - \frac{a^2}{m^2} - \frac{3r_0}{2m} } \right), \label{eq:horiz}
\end{equation}
where $k = 1$ indicates the smaller radius, while $k = 2$ indicates the largest radius, and the negative roots were ignored because the radial coordinate is assumed to be
positive.

By inspecting the expressions for the roots just given, and taking into account that the free parameters $m/r_0$ and $a/r_0$ are nonnegative,  we can see that $r_{1-}$ is complex for all nonvanishing values of the free parameters. Hence, the interior solution can have at most one horizon. This happens if the root $r_{2-}$ is such that $r_{2-}\le r_0$. Moreover, the complete solution can have at most two horizons. To see this, let us consider the limiting case that arises under the equality of the horizon functions, i.e., $\Delta_-(r= r_0)= \Delta_+(r = r_0) = 0$. This equality yields the limiting mass
\begin{align}
    \frac{m_l}{r_0} = 2\left(1 + \frac{a^2}{r_0^2}\right)\label{eq:ml1},
\end{align}
from which we can draw several properties of the resulting geometries. 

For masses larger than $m_l$, i.e., for $m/r_0 >m_l/r_0$, the inner Cauchy horizon is located in the interior matter region and the exterior (event) horizon is located outside matter. 

For $m/r_0 = m_l/r_0$, the inner Cauchy horizon is located at the boundary of the matter distribution, and the exterior (event) horizon is located outside the matter.

\begin{figure}
    \centering
    \includegraphics[width=.45\columnwidth]{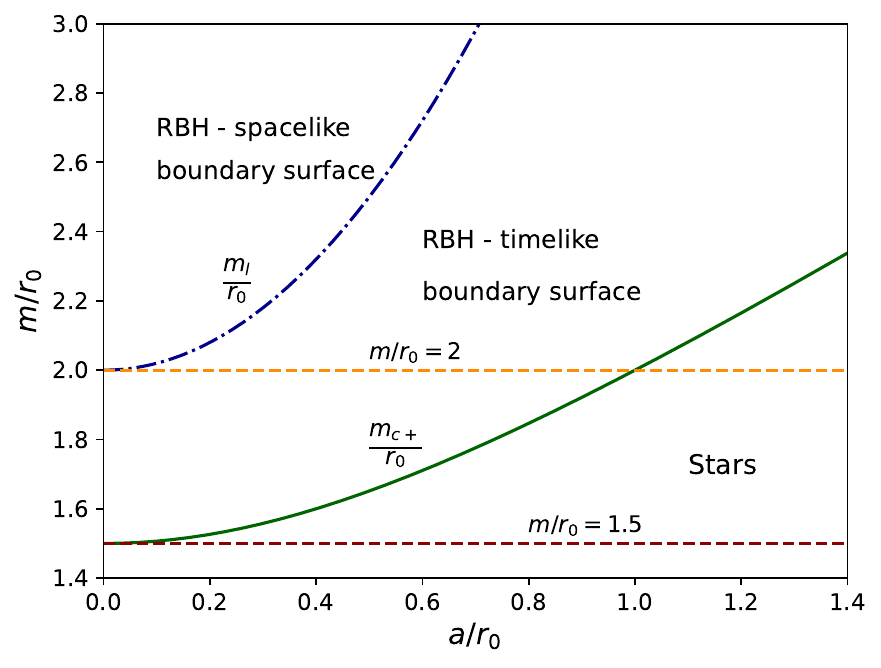}
    \caption{A plot showing the limiting and extremal masses as a function of $a/r_0$. The label on each curve indicates the respective normalized mass. Such mass functions are needed in order to uncover the different kinds of objects and the properties of the boundary surfaces. The horizontal lines $m/r_0 = 2$ and $m/r_0 = 1.5$ are important to establish the existence and location of ergosurfaces.}
    \label{fig:mr0}
\end{figure}

For masses smaller than $m_l$, i.e., for $m/r_0 < m_l/r_0$, the complete solution can present two, one, or no horizons. When presenting horizons, they are located outside the matter distribution. Here, the extremal case that separates spacetimes with no horizon from spacetimes with two horizons is the case with a degenerate horizon, which is equivalent to the vanishing of the discriminant of $\Delta_+(r)$, and that yields the extremal mass $m_{c+}$
\begin{align}
    \frac{m_{c+}}{r_0}  = \frac{3}{4}\left(1 + \sqrt{1 + \frac{16}{9}\frac{a^2}{r_0^2}} \right) \label{eq:discri3}.
\end{align}
The behavior of this extremal mass and the limiting mass \eqref{eq:ml1} as a function of the rotation parameter $a/r_0$ are shown in Fig.~\ref{fig:mr0}.

Now we turn our attention to the ergoregions. 
The locations of the ergosurfaces that arise in the rotating version of the L\&Z charged solution can also be obtained analytically, and are given by similar expressions as Eq.~\eqref{eq:horiz}, just by replacing $a^2$ with $a^2 \cos^2 \theta$, which yields
\begin{equation}
   r_{k-}(\theta) = r_0 \sqrt{\frac{r_0}{m}\left(1 + (-1)^k \sqrt{1 +  \frac{2m}{r_0} \frac{a^2\cos^2\theta}{r_0^2}}\right)}, \qquad \  r_{k+}(\theta) = m \left(1 + (-1)^k \sqrt{1 - \frac{a^2\cos^2\theta}{m^2} - \frac{3r_0}{2m}} \right). \label{eq:ergo}
\end{equation}
The situation for the ergosurface solutions $r_{k\mp}(\theta)$ is similar to that for the horizons $r_{k\mp}$.

Considering the locations of horizons \eqref{eq:horiz} and ergosurfaces \eqref{eq:ergo}, we may draw a set of statements regarding the different types of objects that the present geometry allows. The main classes of objects are the same as those described in our related work~\cite{BZ2024-paper1}, and then we give just a brief description of them below. Figure \ref{fig:mr0} summarizes the situation here.  

In the interval of large masses, characterized by $m/r_0 > m_l/r_0$ and corresponding the region above the dashed-dotted line $m/r_0 = m_l/r_0$ in Fig.~\ref{fig:mr0}, the complete solution presents two horizons: an inner Cauchy horizon situated inside matter, and an exterior (event) horizon located outside matter. This kind of configurations represent charged rotating nonextremal RBH with a spacelike boundary. 

In the critical limit of large masses, i.e., for  $m/r_0 = m_l/r_0$, 
the solution exhibits two horizons: an inner Cauchy horizon situated at the boundary of the matter distribution, and an exterior event horizon located outside the matter. This solution represents a charged rotating nonextremal RBH with a lightlike boundary. Such a configuration can be viewed as the limit of cases with a timelike boundary. Alternatively, it can be derived more rigorously by imposing smooth junction conditions on a null hypersurface, following the procedure outlined in Appendix A of Ref.~\cite{Masa22}. 

In the interval of intermediate masses, characterized by $m_{c+}/r_0 < m/r_0 < m_l/r_0$ and corresponding to the region delimited by the solid line $m/r_0 = m_{c+}/r_0$ and the dashed-dotted line $m/r_0 = m_l/r_0$ in Fig.~\ref{fig:mr0}, the solutions present two horizons, both located outside matter. The corresponding configurations represent charged rotating nonextremal RBH with a timelike boundary.

In the lower limit of intermediate masses, i.e., for $m/r_0 = m_{c+}/r_0$, the solution presents a degenerate horizon outside matter, and represents a charged rotating extremal RBH with a timelike boundary. 

Finally, in the interval of small masses, characterized by masses in the interval $m/r_0 < m_{c+}/r_0$ and corresponding to the region below the $m/r_0 = m_{c+}/r_0$ in Fig.~\ref{fig:mr0}, the solutions present no horizons, and the corresponding configurations represent charged rotating overextremal star-like configurations with a timelike boundary.

Regarding the existence and location of ergosurfaces, we may draw the following statements. 

Configurations characterized by parameters belonging to the regions above the dashed line $m/r_0 = 2$ in Fig.~\ref{fig:mr0} exhibit two distinct ergosurfaces. That region includes charged rotating RBH with a spacelike boundary, charged rotating RBH with a lightlike boundary, charged rotating RBH with a timelike boundary, and star-like objects. 
For charged rotating non-extremal RBHs featuring either a spacelike or lightlike boundary, the inner ergosurface lies entirely within the matter distribution, whereas the outer ergosurface is located completely outside matter. In both non-extremal and extremal charged rotating RBHs with a timelike boundary, the inner ergosurface extends across the matter distribution, whereas the outer ergosurface is located entirely in the exterior region. In configurations representing star-like objects, the two ergosurfaces extend from the interior spacetime into the exterior region.  

Configurations characterized by parameters within the region between the dashed lines $m/r_0 = 1.5$ and $m/r_0 = 2$ in Fig.~\ref{fig:mr0}, which encompasses charged rotating nonextremal and extremal RBH with timelike boundaries, as well as star-like objects, exhibit two ergosurfaces, both of which lie entirely outside the matter distribution.

Finally, configurations characterized by parameters in the region below the dashed line $m/r_0 = 1.5$ in Fig.~\ref{fig:mr0}, which encompasses stars-like objects, do not exhibit ergosurfaces. This is consistent with the typical behavior of overcharged rotating objects in Einstein-Maxwell theory.

\section{An alternative interior electromagnetic field for the rotating electrically charged solution}
\label{sec:altEM}

\subsection{The geometry}

As previously noted, the same global geometry can accommodate different interior electromagnetic fields. 
It is therefore of interest to explore alternative configurations for the interior electromagnetic field while maintaining interior and exterior metrics identical to those of the previous solution. 
The interior geometry is given by the G\"urses-G\"ursey metric~\eqref{eq:ggmetric} with the mass function $M(r)$ given by Eq.~\eqref{eq:m-(r)}. 
Such an interior metric is matched to the standard Kerr-Newman metric, which is recovered from \eqref{eq:ggmetric} when $\mu(r)=m$ and $Q(r)=q$ are constant, such that $M(r)=m - q^2/(2 r)$.

\subsection{The interior electromagnetic field}
\label{sec:intEMb}

 Following the same procedure as described in Sec.~II A of Ref.~\cite{BZ2024-paper1}, we may choose an alternative interior solution for the electromagnetic gauge potential in comparison to the solution presented in Sec.~\ref{sec:rotLZ}. A suitable choice is 
\begin{align}
    \mathcal{A}_{\mu} = \frac{q}{r_0}\frac{r^2}{\Sigma}\left(\delta_{\mu}^{\ t} - a \sin^2 \theta \delta_{\mu}^{\ \varphi}\right). \label{eq:rotpot} 
\end{align} 
This gauge potential has the same form as the electromagnetic gauge potential of the Kerr-Newman solution, but here with the constant charge $q$ replaced by the function $Q(r) = \dfrac{q\, r}{r_0}$. 
Notice also that, in the limit of the vanishing rotation parameter $a=0$, the interior gauge potential \eqref{eq:rotpot} reduces to $ {\cal A}_\mu= \dfrac{q}{r_0}\,\delta_{\mu}^{\ t}$, exactly the interior gauge potential of the static L\&Z solution.

In the present case, the nontrivial components of the Faraday-Maxwell tensor simplify to
\begin{equation}
\begin{aligned}
   & F_{rt} = -\frac{2q r a^2 \cos^2 \theta}{r_0 \Sigma^2}, \qquad \ F_{r\varphi} = - a \sin^2 \theta F_{ rt},  \qquad F_{\theta t} = - \frac{qr^2a^2 \sin 2 \theta}{r_0\Sigma^2}, \qquad \  F_{\theta \varphi} = -\frac{\left(r^2+a^2\right)}{a} F_{\theta t}, \label{eq:electenLZ1}
\end{aligned}
\end{equation}
which all vanish in the limit $a \to 0$, recovering the static interior solution. 
It can also be seen from the result \eqref{eq:electenLZ1} that the Faraday-Maxwell tensor is well-behaved in the entire interior region, including the ring $S^1$.

From the relations and Eq.~\eqref{eq:ErBr}, we can see that the electric and magnetic fields, as measured by the comoving observer, are parallel to each other and given by
\begin{align}
     E_r = - \sqrt{\frac{\Sigma}{\pm \Delta}} \frac{2q r a^2 \cos^2 \theta}{r_0 \Sigma^2}, \qquad \  B_r = \sqrt{\frac{\Sigma}{\pm \Delta}} \frac{2 qr^2a \cos \theta}{r_0\Sigma^2}.
\end{align}
In this alternative solution for the interior electromagnetic field, the electric field does not vanish. Hence, in this case the medium does not correspond to a perfect conductor. 

The current density generated by the Faraday-Maxwell field \eqref{eq:electenLZ} is given by
\begin{align}
    4\pi J^t =  \frac{2q a^2}{r_0} \left(\frac{r^2 \sin^2 \theta-(r^2 + a^2)\cos^2\theta}{\Sigma^3} \right), \qquad\ \  4 \pi J^{\varphi} = \frac{2qa}{r_0}\left(\frac{r^2-a^2\cos^2\theta}{\Sigma^3} \right), \label{eq:Jphi-1}
\end{align}
with the other components being identically zero. Notice that $J^t$ is positive for $\tan^2 \theta > (r^2 +a^2)/a^2$ and is negative for $\tan^2 \theta < (r^2 +a^2)/a^2$, and then we can interpret the interior fluid as an electrically polarized medium. Moreover, as seen from Eq.~\eqref{eq:Jphi-1}, the two nontrivial components of the current density are finite everywhere in the interior region, with the exception of the ring $S^1$, where both components diverge. Inside the ring, on the disc $(r=0,\, 0\leq \theta<\pi/2)$,  the two components of the current density are finite and are given by $J^t = -2q/r_0 a^2 \cos^4 \theta$ and $J^{\varphi} = -2q/r_0 a^3 \cos^4 \theta$, respectively. On the other hand, the Faraday-Maxwell tensor given by~\eqref{eq:electenLZ1} is finite everywhere inside the fluid and vanishes at the ring, and on the central disk.

The decomposition of the current density $J^{\mu}$ along the Carter tetrad, given by Eqs.~\eqref{eq:jmu-} and~\eqref{eq:Jdecomp}, allows us to obtain
\begin{align}
    \rho_{e} = -\sqrt{\frac{\pm \Delta}{\Sigma}} \frac{q  a^2 \cos^2 \theta}{2 \pi r_0 \Sigma^2}, \qquad \ \ j^{\mu} = \frac{qr^2}{2 \pi r_0 \Sigma^2} \frac{a \sin \theta}{\sqrt{\Sigma}} e_3^{\ \mu}, \label{eq:conduction-1}
\end{align}
where $ e_3^{\ \mu}$ is the fourth component of the Carter tetrad. Since the interior medium does not correspond to a perfect conductor and the conduction current $j^\mu$ does not vanish, we can associate the internal electric field to the conduction current through the following well-known relation (see, e.g., \cite{Bekenstein78})
\begin{align}
    j^{\mu} = \kappa^{\mu \nu} E_{\nu},
\end{align}
where $\kappa^{\mu \nu}$ is the symmetrical conductivity tensor which, in the present case, is given by
\begin{align}
    \kappa^{\mu \nu} = - \frac{r \sin \theta}{4 \pi a  \sqrt{\Sigma}\cos^2 \theta} e_{1}^{\ \mu} e_{3}^{\ \nu}.
\end{align}
Hence, given that the conduction current does not flow in the same direction as the electric field, this medium may be interpreted as a non-isotropic conductor.

At last, we calculate the net electric charge in the interior of the spheroidal surface of constant radius $r$. Equation \eqref{eq:totalq} furnishes
 \begin{align}
     \mathfrak{q}(r) = q - \frac{q}{r_0} \frac{r^2+a^2}{a} \arctan \frac{a}{r}, \label{eq:totalqq}
 \end{align}
where the last term of this relation carries the whole contribution due to rotation.

\subsection{The exterior electromagnetic field: Kerr-Newman}
As initially assumed, independent of the interior electromagnetic field, the exterior solution is the well-known Kerr-Newman electromagnetic field, whose
gauge potential and Faraday-Maxwell tensor are the same as in the case studied in Sec.~\ref{sec:rotLZ}, given by Eqs.~\eqref{eq:rotpot+} and~\eqref{eq:electen+}, respectively.

\subsection{Matching conditions for the electromagnetic field}
\label{sec:match-eletr}

The usual matching conditions for the electromagnetic fields are given by $[A_a] =0$, $[F_{ab}] = 0$ and $[F_{an}] = 4 \pi \sigma_e u_a$, where $\sigma_e$ is the surface charge density of the shell and $u^a$ is the four-velocity of the boundary, which is given by $0-$element of the tetrad in Eq.~\eqref{eq:rtetrad}. It is straightforward to verify that the first two conditions are trivially satisfied, while the last condition gives us the charge density of the shell as measured by an observer comoving with the boundary, i.e.,
\begin{align}
    \sigma_e= \frac{q}{4 \pi \left(r_0^2 + a^2 \cos^2 \theta\right)}. \label{eq:rotchdlz1}
\end{align}
Notice that the result for the static solution \eqref{eq:chargeden} is recovered in the limit $a=0$. 

The integration of the charge density \eqref{eq:rotchdlz1} over the surface $r=r_0$ gives us
\begin{align}
    \mathfrak{q}_s(r_0) = \frac{q}{r_0} \frac{r_0^2+a^2}{a} \arctan \frac{a}{r_0}. \label{eq:surfq}
\end{align}
Thus, combining Eqs.~\eqref{eq:totalqq} and~\eqref{eq:surfq}, we can see that the total electric charge of the complete solution is $q$, i.e., the total charge inside and on the surface $r=r_0$ is $\mathfrak{q}(r_0) + \mathfrak{q}_s(r_0) = q$. Hence, in contrast to the case discussed in Sec.~\ref{sec:rotLZ}, the interior region of the shell also contributes to the total electric charge of the fluid distribution.

\section{Discussion}
\label{sec:conc}
 
In this work, we derived and analyzed a novel interior rotating geometry matched to the exterior Kerr-Newman solution. Utilizing the Gürses-Gürsey algorithm, we obtained the metric functions and electromagnetic fields as exact solutions to the Einstein-Maxwell equations. By employing the Lemos-Zanchin (L\&Z) static charged metric --which describes charged regular black hole-- as the seed, we generated a rotating counterpart. The resulting configuration features a Kerr-Newman exterior coupled to a rotating de Sitter core, with a rotating electrically charged shell mediating the junction between the two spacetime regions.

In pursuing this goal, we established that for a specific class of electromagnetic fields, a well-behaved interior field cannot exist without a non-zero current density, unless the fields vanish entirely. Unlike the static L\&Z solution, a total absence of the interior electromagnetic field results in a violation of the boundary conditions.

Furthermore, we demonstrated that while the overall electric charge of the static system is preserved, the flexibility of the Gürses-Gürsey procedure allows for the construction of diverse Faraday-Maxwell tensors. This flexibility enables the existence of varied electromagnetic fields and distinct charge densities within the rotating shell, all of which remain compatible with the Kerr-Newman exterior and the same interior/exterior metrics. Specifically, we have constructed two unique interior electromagnetic fields that yield differing current densities within the rotating shell.

In the first case, reported in Sec.~\ref{sec:rotLZ}, the gauge potential and the Faraday–Maxwell tensor depend only on the polar coordinate $\theta$, giving rise to a nonvanishing current density within the shell. Despite this, the net charge enclosed by any spheroidal surface with constant $r<r_0$ vanishes, and the total electric charge is entirely concentrated on the boundary shell. The electromagnetic fields remain finite throughout the entire interior region, including the ring, whereas the current density diverges only at the ring. We interpret the presence of a nonzero interior current density as indicating that the fluid becomes electrically polarized due to the charged rotating spheroidal shell, with the internal medium effectively behaving as a perfect conductor.

In the second case, described in Sec.~\ref{sec:altEM}, the gauge potential and the Faraday–Maxwell tensor depend on both coordinates $r$ and $\theta$, again leading to a nonvanishing current density within the shell. In this configuration, the net electric charge contained within the interior of any spheroid of constant radial coordinate $r<r_0$ is nonzero and varies with $r$. The boundary shell also contributes to the electric charge, ensuring that the total charge of the geometry coincides with that of the static seed solution. As in the previous case, the electromagnetic fields are finite everywhere in the interior spacetime, including at the ring, while the divergence of the current density is confined solely to the ring. We interpret this situation as one in which the total electric charge originates from the interior region, with the rotating fluid being electrically polarized and behaving as a non-isotropic conductor.

Furthermore, we performed a comprehensive analysis of the charged rotating geometry, examining its curvature regularity, energy-momentum tensor, and energy conditions. The investigation revealed a diverse spectrum of configurations, including charged rotating regular black holes with spacelike, lightlike, or timelike boundaries, as well as overextremal regular star-like objects characterized by a timelike boundary.

A natural extension of the present study is to investigate whether interior electromagnetic field configurations not aligned with the principal null directions of the spacetime can be constructed, while still matching the Kerr–Newman electromagnetic field in the exterior region. At present, it remains unclear whether such configurations are physically realizable. In addition, a stability analysis of the regular compact objects presented here is deferred to future work.

\section*{Acknowledgments}
We thank J.~P.~S.~Lemos for stimulating discussions.
MLWB is funded by Funda\c c\~ao de Amparo \`a Pesquisa do Estado de S\~ao Paulo (FAPESP), Brazil, Grant No.~2022/09496-8 and Grant No.~2025/07325-0. VTZ is partly funded by Con\-se\-lho Nacional de Desenvolvimento Cien\-t\'ifico e Tecnol\'ogico (CNPq), Brazil, Grant No.~311726/2022-4.  VTZ thanks Center for Astrophysics and Gravitation (CENTRA),  Instituto Superior T\'ecnico, for hospitality.  

\appendix 
\section{Basic equations of the rotating charged geometry}
\label{sec:basequ}

For completeness, here we provide the main equations that are needed to obtain the metric and electromagnetic functions presented in the main text. The starting point is the Einstein-Maxwell equations, which are expressed as follows,
\begin{align}
    & G_{\mu \nu} =  8\pi T_{\mu \nu} = 8\pi\left( E_{\mu \nu} + M_{\mu \nu}\right),\label{eq:Einst}\\
    &\nabla_{\nu} F^{\mu \nu} = 4 \pi J^{\mu}, \label{eq:Maxw}\\
    & \nabla_{\alpha} F_{\mu \nu} + \nabla_{\nu} F_{\alpha \mu} + \nabla_{\mu} F_{\nu \alpha} = 0. \label{eq:Max2} 
\end{align}
Here, the Greek indices range from $0$ to $3$. The left-hand side of Eq.~\eqref{eq:Einst} consists of the Einstein tensor $G_{\mu \nu}= R_{\mu \nu}- \frac{1}{2}g_{\mu \nu} \mathcal R$, where $R_{\mu\nu}$ denotes the Ricci tensor, $g_{\mu \nu}$ represents the metric tensor, and $\mathcal R$ stands for the Ricci scalar. Equations~\eqref{eq:Maxw} and~\eqref{eq:Max2} involve the covariant derivative of the Faraday-Maxwell tensor $F^{\mu\nu}$ and the current density $J^\mu$. As is well known, Eq. \eqref{eq:Max2} is solved by  expressing the Faraday-Maxwell tensor in terms of the electromagnetic gauge potential $\mathcal{A}_{\mu}$ as $F_{\mu \nu} = \nabla_{\nu} \mathcal{A}_{\mu} - \nabla_{\mu} \mathcal{A}_{\nu}$, with $\nabla_{\mu}$ representing the covariant derivative compatible with the four-dimensional Lorentzian metric.  

The energy-momentum tensor (EMT) $T_{\mu \nu}$ has two components and can be expressed as $T_{\mu \nu} = E_{\mu \nu} + M_{\mu \nu}$. The former stems from the electromagnetic field, while the latter originates from the matter itself. The electromagnetic EMT $E_{\mu \nu} $ takes the form
\begin{align} 
    E_{\mu \nu} = \frac{1}{4\pi}\left( F_{\mu \alpha} F_{\nu}^{\ \alpha} - \frac{1}{4}g_{\mu \nu} F_{\alpha \beta}F^{\alpha \beta}\right).
\end{align}
Meanwhile, the form of the matter energy-momentum tensor (EMT) $M_{\mu \nu}$ depends on whether the object is rotating or not. In the static case, where the matter content is a charged perfect fluid, the EMT takes the standard form $M_{\mu\nu}= \left(\rho_m+p\right)u_\mu u_\nu + p\,g_{\mu\nu}$. However, in the rotating case, the form of the EMT is more complex, as rotation breaks isotropy. In general, the EMT of a rotating charged fluid can be written as $M_{\mu\nu}=\left(\rho_m+p\right)u_\mu u_\nu + p\,g_{\mu\nu}+ \sigma_{\mu\nu}$, where $\sigma_{\mu \nu}$ encodes the anisotropic stresses introduced by rotation.

The effective EMT can be diagonalized through the Carter's orthonormal tetrad $e_i^{\ \mu}$ ($i=0,\,1,\, 2,\,3$), which is given by 
\begin{align}
    & e_0^{\ \mu} = \frac{1}{\sqrt{\pm \Delta \Sigma}}\big[\left(r^2 +a^2\right)\delta^{\mu}_{\ t} + a\, \delta^{\mu}_{\ \varphi}\big], \quad
     e_1^{\ \mu} = \sqrt{\frac{\pm \Delta}{\Sigma}} \delta^{\mu}_{\ r}, \quad  e_2^{\ \mu} = \frac{1}{\sqrt{\Sigma}} \delta^{\mu}_{\ \theta},
    \quad e_3^{\ \mu} =  \frac{1}{\sqrt{\Sigma}\sin \theta}\big[a \sin^2 \theta\delta^{\mu}_{\ t} + \delta^{\mu}_{\ \varphi}\big]. \label{eq:rtetrad} 
\end{align}
The signs  $\pm$ in front of $\Delta$ apply to the different spacetime regions where $\Delta(r)$ is positive and negative, respectively. 
That is, the plus sign holds in regions outside the event horizon and inside the Cauchy horizon, where $\Delta$ is positive and $e_0^{\ \mu}$ is a timelike vector that represents the four-velocity of a stationary observer with angular velocity $\Omega(r) = e_0^{\ \varphi}/e_0^{\ t} = a/(r^2 + a^2)$, while $e_1^{\ \mu}$ is a spacelike vector. The minus sign holds in the region between the event horizon and the Cauchy horizon, where $\Delta(r)$ is negative, and the vectors $e_0^{\ \mu}$ and $e_1^{\ \mu}$ exchange roles. Therefore, in this region, $e_1^{\ \mu}$ can be interpreted as the four-velocity of the comoving observer with the fluid. 
Finally, the elements of the tetrad $e_2^{\ \mu}$ and $e_3^{\ \mu}$ are spacelike in all regions.

In the tetrad base \eqref{eq:rtetrad}, the effective EMT takes the form~\cite{Gurses, Burinskii, Gondolo}
\begin{align}
    T^{\mu \nu} = \varrho\, e_0^{\ \mu}e_0^{\ \nu} + \mathfrak p_1 e_1^{\ \mu}e_1^{\ \nu} + \mathfrak p_{2} e_2^{\ \mu}e_2^{\ \nu} + \mathfrak p_{3} e_3^{\ \mu}e_3^{\ \nu},\label{eq:ggt}
\end{align}
where $\varrho\equiv\varrho(r,\theta)$, $\mathfrak p_{1}\equiv\mathfrak p_1(r,\theta)$, $\mathfrak p_{2}\equiv\mathfrak p_2(r,\theta)$, and $\mathfrak p_{2}\equiv \mathfrak p_3(r,\theta)$ are given by (see, e.g., \cite{Burinskii})
\begin{align}
    & \varrho(r,\theta) = -\mathfrak p_1(r,\theta) = \frac{r^2 M'}{4 \pi \Sigma^2}, \ \ \ \mathfrak p_{2}(r,\theta) = \mathfrak p_{3}(r,\theta) = \frac{r^2 M'}{4 \pi \Sigma^2} - \frac{1}{8 \pi \Sigma}\Big(r\,M\Big)''. \label{eq:tpressa}
\end{align}
 
The projection of the electromagnetic EMT onto the Carter's orthonormal frame yields the following nonvanishing components,
\begin{equation} \label{eq:emtEMa}
\begin{aligned}
    & \varrho_{em}(r, \theta) \equiv E_{\mu \nu}e_{0}^{\ \mu} e_{0}^{\ \nu} = \frac{1}{8 \pi}\left(F^2_{rt} + \frac{F^2_{\theta t}}{a^2 \sin^2 \theta} \right), \qquad 
     & \mathfrak p_{em 1}(r,\theta) \equiv E_{\mu \nu}e_{1}^{\ \mu} e_{1}^{\ \nu} = - \varrho_{em}(r, \theta),\\
    & \mathfrak p_{em 2}(r,\theta) \equiv E_{\mu \nu}e_{2}^{\ \mu} e_{2}^{\ \nu} = \varrho_{em}(r, \theta), \qquad 
     &\mathfrak p_{em 3}(r,\theta) \equiv  E_{\mu \nu}e_{3}^{\ \mu} e_{3}^{\ \nu} = \varrho_{em}(r, \theta). 
\end{aligned}
\end{equation}
It is worth noting that when $e_0^{\ \mu}$ and $e_1^{\ \mu}$ swap roles, a corresponding interchange occurs between the effective energy density $\varrho$ and the radial pressure $\mathfrak p_1$. This symmetry extends to the corresponding electromagnetic quantities $\varrho_{em}$ $\mathfrak p_{em 1}$ as well.

Then, it is easy to see that the EMT of the matter can be determined by the difference between the total EMT and the electromagnetic EMT,  
i.e., $M_{\mu \nu} e_{a}^{\ \mu} e_{b}^{\ \nu} = (T_{\mu \nu} - E_{\mu \nu})e_{a}^{\ \mu} e_{b}^{\ \nu}$, and it corresponds to the EMT of a non-isotropic fluid whose energy density and pressures are give by
\begin{equation}\label{eq:emt-matter}
\begin{aligned}
    & \varrho_{m}(r, \theta) \equiv M_{\mu \nu} e_{0}^{\ \mu}e_{0}^{\ \nu}= \varrho(r,\theta) - \varrho_{em}(r,\theta) ,\qquad \
    & \mathfrak p_{m 1}(r,\theta) \equiv M_{\mu \nu} e_{1}^{\ \mu} e_{1}^{\ \nu} = -\varrho(r,\theta) +\varrho_{em}(r,\theta),\\
    & \mathfrak p_{m 2}(r,\theta) \equiv M_{\mu \nu} e_{2}^{\ \mu} e_{2}^{\ \nu}=\mathfrak p_2(r,\theta) - \varrho_{em}(r,\theta), \qquad 
    & \mathfrak p_{m 3}(r,\theta) \equiv M_{\mu \nu} e_{3}^{\ \mu} e_{3}^{\ \nu}=\mathfrak p_2(r,\theta) - \varrho_{em}(r,\theta).
\end{aligned}
\end{equation}
These relations are used in the main text.

\section{G\"urses-G\"ursey metric and the Maxwell equations}
\label{sec:appMaxwell}

\subsection{The Maxwell equations}
Here we present some more details of the derivation of the Maxwell equations presented in the main text. We also calculate the resulting current density for the two choices of electromagnetic fields analyzed in the text. The set of Maxwell equations can be written in the form 
\begin{align}
    & \partial_r\left[\left(r^2 +a^2\right)\sin \theta F_{rt}\right] + \partial_{\theta}\left[\sin \theta F_{\theta t}\right] = 4 \pi \sqrt{-g} J^t, \ \ \ \partial_r \left[\csc\theta F_{r \varphi} \right] + \partial_{\theta} \left[\frac{\csc\theta}{r^2 + a^2} F_{\theta \varphi} \right] = - 4 \pi \sqrt{-g} J^{\varphi}, \label{eq:mx2} \\
    & \partial_{r} F_{\theta t} -  \partial_{\theta} F_{rt} = 0, \ \ \ \ \ \ \ \ \ \ \ \ \ \ \ \ \ \ \ \ \ \ \ \ \ \ \ \ \ \ \ \ \ \ \ \ \ \ \ \ \ \ \ \partial_{r} F_{\theta \varphi} - \partial_{\theta} F_{r \varphi} = 0, \label{eq:mx4}
\end{align}
where $\partial_{\mu} \equiv \partial/\partial x^{\mu}$, $g = - \Sigma^2 \sin^2 \theta$ is the determinant of the metric~\eqref{eq:ggmetric}, and $J^{\mu}$ is the current density. The set of  Eqs.~\eqref{eq:mx2} determine the current density inside the charged matter that generates the electromagnetic field, while the set of Eqs.~\eqref{eq:mx4} are the compatibility equations. 

It is worth noting that the Maxwell equations in the G\"urses-G\"ursey geometry are independent of the mass function $M(r)$. This means that the solutions of such equations are determined once we give the sources $J^t(r,\theta)$ and $J^\varphi(r,\theta)$. Conversely,  one can make an ansatz for the gauge potential, or for the Faraday-Maxwell tensor, and determine the current density from the Maxwell equations.

\subsection{Interior electromagnetic field of Sec.~\ref{sec:intEMa} of the main text}

The electromagnetic gauge potential is given by Eq.~\eqref{eq:rotpot1}, from which it results that the interior Farady-Maxwell tensor $F_{\mu\nu}$ is given by Eqs.~\eqref{eq:Frt0} and \eqref{eq:electenLZ}.

Now, it is straightforward to show that the Maxwell equations~\eqref{eq:mx2}--\eqref{eq:mx4} are indeed satisfied for the interior electromagnetic gauge potential \eqref{eq:rotpot1}.
First, since $F_{rt}$ and $ F_{r \varphi}$ vanish and $F_{\theta t}$ and $F_{\theta \varphi}$ are only functions of $\theta$, cf. Eqs.~\eqref{eq:Frt0} and~\eqref{eq:electenLZ}, it is easy to see that the equations in~\eqref{eq:mx4} are trivially satisfied. In turn, the first set of Eqs.~\eqref{eq:mx2} are also satisfied for the current density given by Eq.~\eqref{eq:Jphi-}. In fact, these two equations provide the nontrivial components of the current density $J^\mu$ in \eqref{eq:Jphi-}.

\subsection{Interior electromagnetic field of Sec.~\ref{sec:intEMb} of the main text}

To begin with, observe that the non-vanishing components of the electromagnetic tensor given in Eq.~\eqref{eq:electenLZ1} are connected through the relations in Eq.~\eqref{eq:a1}. This relation enables us to reformulate the Maxwell equations~\eqref{eq:mx2}--\eqref{eq:mx4} exclusively in terms of the components $F_{rt}$ and $F_{\theta t}$, namely,
\begin{align}
    & \partial_r\left[\left(r^2 +a^2\right)\sin \theta F_{rt}\right] + \partial_{\theta}\left[\sin \theta F_{\theta t}\right] = 4 \pi \sqrt{-g}J^t, \qquad \partial_r \left[a \sin \theta F_{r t} \right] + \partial_{\theta} \left[\frac{\csc\theta}{a} F_{\theta t} \right] = 4 \pi\sqrt{-g} J^{\varphi}, \label{eq:mx6ap} \\
    & \partial_{r} F_{\theta t} -  \partial_{\theta} F_{rt} = 0, \qquad \qquad \qquad \qquad \qquad \qquad \qquad \ \ \ \ \partial_{r}\left[\left(r^2+a^2\right) F_{\theta t}\right] - \partial_{\theta}\left[a^2 \sin^2 \theta F_{r t}\right] = 0. \label{eq:mx8ap}
\end{align}

Now we can show that the compatibility equations~\eqref{eq:mx8ap} are indeed satisfied. From Eq.~\eqref{eq:electenLZ1}, it follows the relation
\begin{align}
    \partial_{r} F_{\theta t}  & =  \partial_{\theta} F_{rt}    =  \frac{2 q r a^2 \sin 2\theta}{r_0 \Sigma^3} \Big(r^2 - a^2 \cos^2 \theta\Big), \nonumber
\end{align}
which implies that first compatibility condition of Eq.~\eqref{eq:mx8ap} is identically satisfied.
The second compatibility condition in Eq.~\eqref{eq:mx8ap}, after using Eq.~\eqref{eq:a1}, can be rewritten in the form
\begin{align}
    \Sigma \partial_{r} F_{\theta t} = a^2 \sin 2 \theta F_{rt} - 2r F_{\theta t},
\end{align}
and then it is an easy task to verify that the electromagnetic field tensor~\eqref{eq:electenLZ1} obeys such a relation. 

The other two Maxwell equations~\eqref{eq:mx6ap}, furnish the nontrivial components of the current density $J^\mu$. Given the components of the electromagnetic tensor in Eq.~\eqref{eq:electenLZ1}, we have
\begin{align}
     \partial_r F_{rt} = -\frac{2q}{r_0} \left(\frac{r^2\left(r^2-3 a^2\cos^2\theta\right)}{\Sigma^3}  + \frac{(r^2-a^2\cos^2\theta)}{\Sigma^2}\right), \ \ \ 
    \partial_{\theta} F_{\theta t} = \, \frac{2qr^2a^2}{r_0\Sigma^3}\Big(\!\!\left[r^2-3 a^2\cos^2\theta\right]\sin^2 \theta - \Sigma \cos^2 \theta \Big).  \nonumber
\end{align}
These two relations allow us to obtain the components of the current density $J^{\mu}$ given by Eq.~\eqref{eq:Jphi-1}.

\end{document}